\documentclass{article}

\usepackage{authblk}
\usepackage{amsmath}
\usepackage{amsthm}
\theoremstyle{definition}

\usepackage{graphicx}
\usepackage{amssymb}
\usepackage{pgfplots}
\usepackage{siunitx}
\usepackage{hyperref}
\usepackage{comment}
\usepackage{subfig}

\begin{document}
\title{Role of the Solar Minimum in the Waiting Time Distribution Throughout the Heliosphere}
\author[1]{Yosia Nurhan}
\author[1]{Jay Johnson}
\author[1]{Jonathan Homan}
\author[1,2]{Simon Wing}
\affil[1]{Andrews University, Berrien Springs, MI, 49103, USA}
\affil[2]{Johns Hopkins University, Baltimore, MD, 21218, USA}
\maketitle

%\collaboration{(LaTeX collaboration)}



\begin{abstract}
We explore the tail of various waiting time datasets of processes that follow a nonstationary Poisson distribution with a sinusoidal driver. Analytically, we find that the distribution of large waiting times of such processes can be described using a power law slope of -2.5. We show that this result applies more broadly to any nonstationary Poisson process driven periodically.  Examples of such processes include solar flares, coronal mass ejections, geomagnetic storms, and substorms. We also discuss how the power law specifically relates to the behavior of driver near its minima.
\end{abstract}

%\keywords{solar minimum, notices --- miscellaneous --- catalogs --- surveys}

%% We recommend that authors also use the natbib \cite

\section{Introduction}
The solar cycle follows an 11-year cycle, characterized by fluctuations in the numbers and surface area of sunspots, and impacts processes at the sun and throughout the heliosphere \cite{Hathaway2010}.   Many processes that occur during the solar cycle can be identified as ``events"  because they are either localized in time or have a well defined onset.
Solar flares, coronal mass ejections (CMEs), geomagnetic storms, and geomagnetic substorms are among this class of processes. 

It is well known that the dynamics of systems can be well described and reconstructed by considering the distribution of the set of time intervals of the system, ($\Delta_1, \Delta_2, \Delta_, ..., \Delta_N $) known as waiting times \cite{Snelling2020, Aschwanden2010, Wheatland2000, Wheatland2002, Wheatland2003, Consolini2002}.%(and many more). 
The waiting time distributions (WTDs) that are observed throughout the heliosphere may be governed by a combination of internal system dynamics and external driving by the magnetic activity cycle of the sun.  If the events occur randomly, the process can be described as a Poisson process characterized by its rate. However, because many events throughout the heliosphere respond directly to the magnetic activity cycle, the activity cycle may modulate the rate of the events. As long as the response is random in nature, such processes can be described in terms of a nonstationary or time dependent Poisson process.

In general, the waiting time for solar flares, CMEs, storms, and substorms can be described by a nonstationary Poisson distribution, especially at large waiting times. 
Using Tsallis statistical mechanics, \cite{Balasis2011a} suggested that the driving physical mechanisms of solar flares and storms have the same characteristics. Recently, \cite{Snelling2020} used information theory to show that while there is a short-term memory in the solar flare sequence, the distribution cannot be distinguished from a nonstationary Poisson distribution for longer waiting times. The WTD of CMEs have also been understood to follow a nonstationary Poisson process \cite{Wheatland2003}. CMEs, along with corotating interaction regions (CIRs), are the main drivers of storms. The CME-driven storms have a random occurrence pattern consistent with a nonstationary Poisson distribution. In contrast, CIR-driven storms have a periodicity of $\sim$27 days. CIR-driven storms usually have a weaker intensity than CME-driven storms and happen during the declining phase of solar maximum \cite{Tsubouchi2007, Borovsky2006}. On average, the WTD of storms follows a nonstationary Poisson process \cite{Tsubouchi2007}. The WTD of substorms have two components: a non-random component which may correspond to spontaneous substorms with characteristic waiting time of $\sim$2.5 hours and a random component which is fit by a Poisson distribution and generally occurs at waiting times longer than 5 hours \cite{Borovsky1993}.

%Flares and magnetic storms time series both follow a  nonextensive energy distribution function, suggesting driving mechanisms that have the same characteristics (https://reader.elsevier.com/reader/sd/pii/S0378437110008290?token=BE2791CEF66A7EED56A6592477BCB67D156E4AF30B85E825C27E5F2EED8AE4663D41D58A6703D592F92DCBFAAA14C455&originRegion=us-east-1&originCreation=20210425232537).

The WTDs of these processes exhibit a power law for longer waiting times (heavy tail). For the WTD of solar flares at long waiting times, \cite{Boffetta1999} found a power law slope of $-2.4 \pm 0.1$. \cite{Wheatland2000}, restricting for flares of class C1 and above, found a power law slope of $-2.16 \pm 0.05$. \cite{Aschwanden2010} found a power law in the tail of the WTD of solar flares with slope of $\sim 2$. \cite{Wheatland2003} found that the distribution of waiting times of CMEs in the Large Angle and Spectrometric Coronagraph (LASCO) CME catalog for the years 1996-2001 exhibit a power law tail of $2.36 \pm 0.11$ for ($\Delta t > 10$ hours). Using $Dst$ index from World Data Center for Geomagnetism, Kyoto University, Japan, \cite{Tsubouchi2007} fitted the tail of the WTD for geomagnetic storms ($Dst$ $< -100$ nT and $\Delta > 48$ hours) for $\Delta t > 1000$ hours with a power law of $~2.2 \pm 0.1$. %\cite{Consolini2002} use the auroral electrojet (AE) index from both the National Geophysical Data Centre (NGDC, Boulder, Colorado) and the World Data Centre for Geomagnetism, Kyoto University Japan from 1 January 1978 to 30 June 1988. They found a 

\begin{figure*}[p] %FIGURE SIZE 7x6 fontsize 12
   \centering
   \includegraphics[width = 15cm]{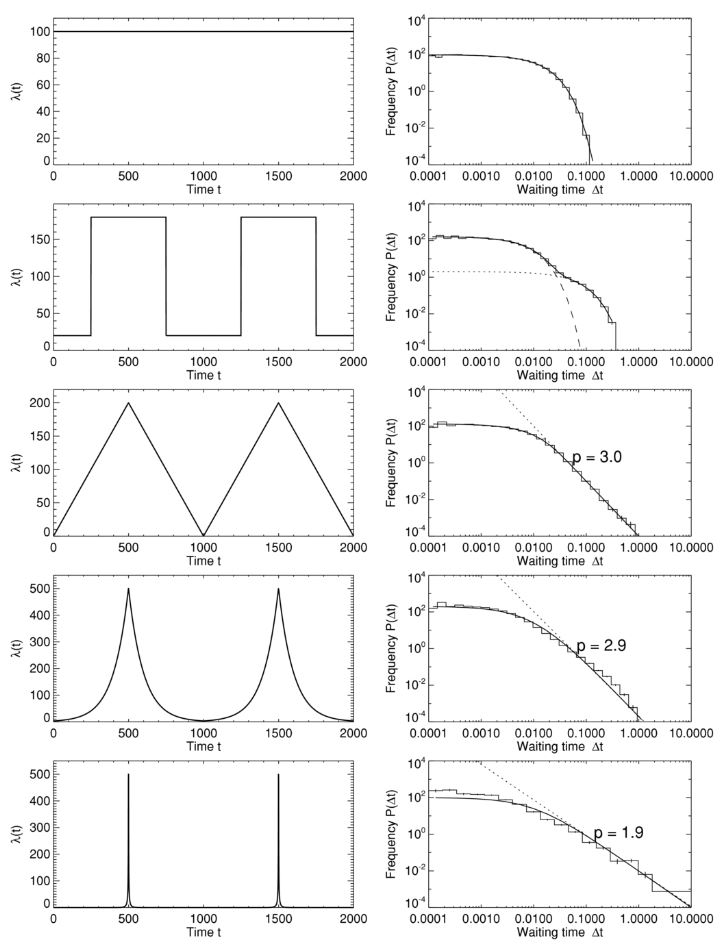}
  \caption{Figure from \cite{Aschwanden2010}. The rate functions, $\lambda(t)$, are shown on the left side, and the corresponding waiting time distributions are shown on the right hand side. Power law fits are indicated with a
dotted line where $p$ is the power law slope.
              }
         \label{fig:asch}
\end{figure*}

Some have proposed different driver mechanisms of the nonstationary Poisson distribution of the WTD of solar flares and explored their consequence to the behavior of the power law slope. 
\cite{Wheatland2003} proposed that the flaring rate of solar flares follows an exponential distribution and showed analytically that the WTDs of such nonstationary Poisson processes follow a power law of -3. \cite{Aschwanden2010} further studied several different drivers and the power law resulted shown in Figure \ref{fig:asch}.

\cite{Wheatland2002} showed that the power law slope of WTDs varies with the solar cycle. Since the solar cycle is approximately sinusoidal, we propose that the flaring rate of solar flares follows the sinusoidal distribution to the first order. Here, we will show analytically how the observed power law behavior originates from the sinusoidal rate and specifically the minima.

\section{Waiting time statistics}

From \cite{Aschwanden2010}, the waiting time probability distribution for a nonstationary Poisson process with continuous flaring rate $\lambda (t)$ is approximately given by the equation
\begin{align}
    P(\Delta) = {\int_0^T \lambda(t)^2 e^{-\lambda(t) \Delta} d t \over \int_0^T \lambda(t) dt}. \label{eq:1}
\end{align}
Suppose we choose a sinusoidal dependence of the flaring rate
\begin{align}
    \lambda(t) = \lambda_0 (1 + \cos \omega t)
\end{align}
for which 
\begin{align}
    \lambda_0 = {1 \over T} \int_0^T \lambda(t) dt .
\end{align}
is the average rate.  Because this is a periodic signal, the statistics can all be obtained considering the
process only over the interval [0 T] where  $T = 2 \pi/\omega$.  Then
\begin{align}
    P(\Delta) = {\lambda_0 e^{-\lambda_0 \Delta} \over T} \int_0^T (1 + \cos(\omega t))^2 e^{-\lambda_0 \cos (\omega t) \Delta} d t 
\end{align}
Changing variables of integration:
\begin{align}
    \theta = \omega t
\end{align}
with 
\begin{align}
    d \theta = \omega d t,
\end{align}
we have
\begin{align}
    P(\Delta) = {\lambda_0 e^{-\lambda_0 \Delta} \over \omega T} \int_0^{2 \pi} (1 + \cos(\theta))^2 e^{-\lambda_0 \cos (\theta) \Delta} d \theta. \label{eq:int}
\end{align}
%Let us consider $P(\Delta)$ determined over a period $\omega T = 2\pi$ (could be more generally $2 \pi n$).
%Then
%\begin{align}
%    P(\Delta) = {\lambda_0 e^{-\lambda_0 \Delta} \over 2 \pi} \int_0^{2 \pi} (1 + \cos\theta)^2 e^{-\lambda_0 \cos \theta \Delta} d \theta \label{eq:int}
%\end{align}
The integral can be performed using the Bessel function identity (\cite{Abramowitz1972} - 9.6.34)
\begin{align}
 e^{z \cos \theta} = \sum_{n = -\infty}^\infty  I_n(z) e^{i n \theta}
\end{align}
 so that  
 \begin{align}
     e^{-\lambda_0 \Delta \cos \theta} = \sum_{n = -\infty}^\infty  I_n(- \lambda_0 \Delta) e^{i n \theta} = \sum_{n = -\infty}^\infty  (-1)^n I_n( \lambda_0 \Delta) e^{i n \theta}
 \end{align}
where          
\begin{align}
    I_n(-z) = (-1)^n I_n(z)  (\text{\cite{Abramowitz1972} - 9.6.30)}.
\end{align}
Then
\begin{align}
    P(\Delta) &= {\lambda_0 e^{-\lambda_0 \Delta} \sum_{n = -\infty}^\infty} (-1)^nI_n(\lambda_0 \Delta) {1 \over 2 \pi} \int_0^{2 \pi} e^{i n \theta}\left(1 + {e^{i \theta} + e^{- i \theta} \over 2}\right)^2  d \theta \\
    &= {\lambda_0 e^{-\lambda_0 \Delta} \sum_{n = -\infty}^\infty} (-1)^nI_n(\lambda_0 \Delta)\left[ {1 \over 2 \pi} \int_0^{2 \pi} e^{i n \theta} \left( 1 + e^{i \theta} + e^{- i \theta}  + {e^{2 i \theta} +  2 + e^{- 2 i \theta} \over 4} \right) d \theta  \right]  \\
    &= {\lambda_0 e^{-\lambda_0 \Delta} \sum_{n = -\infty}^\infty} (-1)^n I_n(\lambda_0 \Delta)\left[{3 \over 2} \delta_{n,0} + \delta_{n,-1} + \delta_{n,1} + {1 \over 4} (\delta_{n,-2} + \delta_{n,2}) \right]  \\
    &= \lambda_0 e^{-\lambda_0 \Delta}  \left[{3 \over 2} I_0(\lambda_0 \Delta) - I_{-1}(\lambda_0 \Delta) - I_1(\lambda_0 \Delta) + {1 \over 4} (I_{-2}(\lambda_0 \Delta) + I_2(\lambda_0 \Delta)) \right]  
\end{align}
and making use of the property
\begin{align}
    I_{-n}(z) &=  I_n(z), \text{(\cite{Abramowitz1972} - 9.6.6)},
\end{align}
we find the analytic solution
\begin{align}
    P(\Delta) &= \lambda_0 e^{-\lambda_0 \Delta}  \left[{3 \over 2} I_0(\lambda_0 \Delta) - 2 I_1(\lambda_0 \Delta) + {1 \over 2}  I_2(\lambda_0 \Delta) \right].  \label{eq:pd}
\end{align}

Now we perform the asymptotic analysis. For large $\Delta$ we have 
\begin{align}
    I_n(z) \sim {e^z \over \sqrt{2 \pi z}}\left[ 1 - {\mu - 1 \over 8 z} + {(\mu - 1) (\mu - 9) \over 2! (8 z)^2} - {(\mu - 1) (\mu - 9) (\mu - 25) \over 3! (8 z)^3} + ... \right]
\end{align}
where $\mu = 4 n^2$. And so
\begin{align}
    I_0(z) e^{- z} &\sim {1 \over \sqrt{2 \pi z}} \left[ 1 + {1 \over 8 z} + {9 \over 128 z^2} + ...\right] \\
    I_1(z) e^{- z} &\sim {1 \over \sqrt{2 \pi z}} \left[ 1 - {3 \over 8 z} - {15 \over 128 z^2} + ...\right]\\
    I_2(z) e^{- z} &\sim {1 \over \sqrt{2 \pi z}} \left[ 1 - {15 \over 8 z} + {105 \over 128 z^2} + ...\right].
\end{align}
Therefore,
\begin{align}
    P(\Delta ) \sim \lambda_0 {1 \over \sqrt{2 \pi \lambda_0 \Delta}} \left[{3 \over 4 (\lambda_0 \Delta)^2} + ...\right] \sim \lambda_0 {3 \over 8} \sqrt{2 \over \pi}  (\lambda_0 \Delta)^{-2.5}.
\end{align}
So, in this case the power law is $-2.5$. In the next section we will show how the power law can be derived from the minima of the rate function, a more general case. 

\begin{figure*}[p]
\centering
\subfloat[WTD]{%
\resizebox*{7cm}{!}{\includegraphics{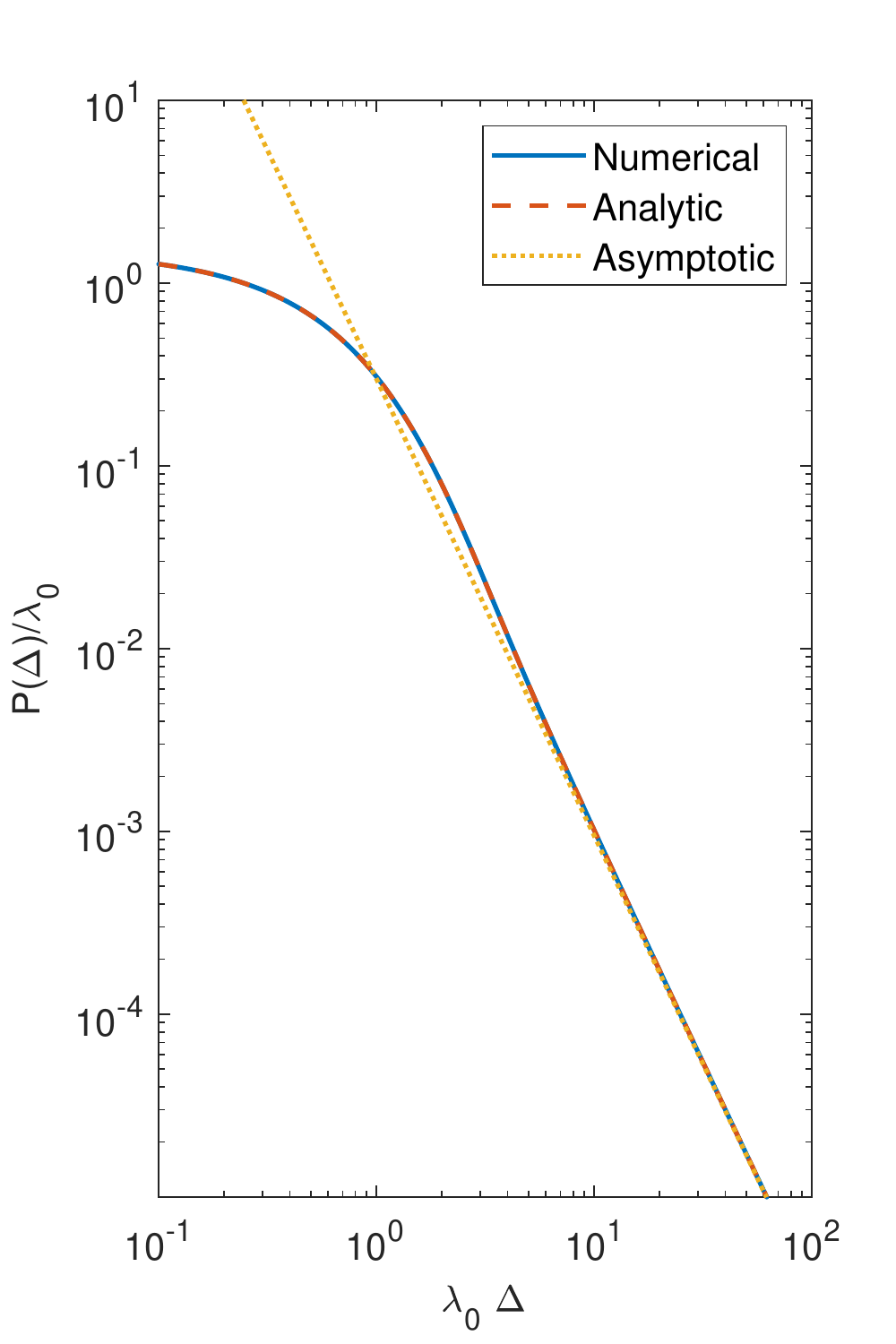}}}\hspace{5pt}
\subfloat[$\alpha$ vs $\lambda_0 \Delta$]{%
\resizebox*{7cm}{!}{\includegraphics{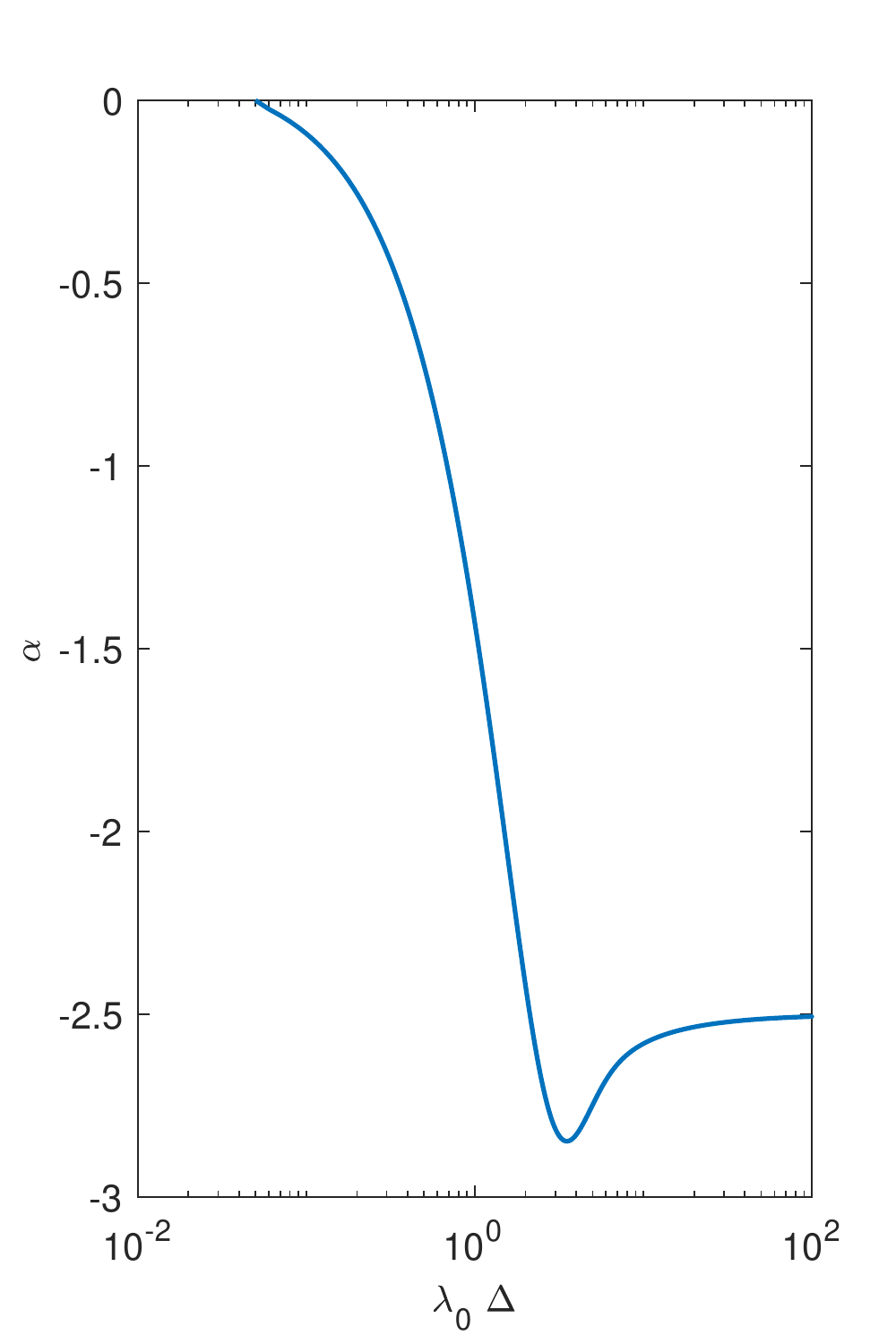}}}
\caption{(a) is a plot of numerical and analytic solutions to eq. (\ref{eq:int}). The Asymptotic solution is $P(\Delta)/\lambda_0 \sim (\lambda_0 \Delta)^{-2.5}$. Plot (b) is the ``local" power law estimation.} 
\label{fig:sim}
\end{figure*}

We confirm our findings by numerically integrating Eq. (\ref{eq:int}). We then plot the numerical and analytic solutions in Figure \ref{fig:sim}. The third plot is the ``local" power slope calculated by numerically differentiating $P(\Delta)/\lambda_0$.  It can be seen that the power law slope approaches -2.5 asymptotically from below. 

\section{Role of the Minima}
The WTD at large waiting times is mainly governed by the minima of the driver. As such, we will show how the minima affects the WTD at large waiting times. As before, from \cite{Aschwanden2010}, for the nonstationary Poisson process with continuous $\lambda (t)$ we have approximately
\begin{align}
    P(\Delta) = \frac{\int_0^T \lambda(t)^2 e^{-\lambda(t) \Delta} d t}{\int_0^T \lambda(t) dt}. \nonumber
\end{align}
Let $\lambda$ be normalized to $\lambda _0$:
\begin{align}
    \lambda = \lambda_0 \cdot g(t).
\end{align}
Then
\begin{align}
    P(\Delta) = \lambda_0{\int_0^T g(t)^2 e^{-\lambda_0\Delta g(t)} d t}. 
\end{align}
Suppose $g(t)$ has a minimum at $t=t_0$:
\begin{align}
    g(t) \approx g(t_0) + \frac{g''(t_0)}{2!}(t-t_0)^2 + ... .
\end{align}
Then, 
\begin{align}
    P(\Delta) = \lambda_0{\int_0^T [g(t_0) + \frac{g''(t_0)}{2!}(t-t_0)^2 + ...]^2 e^{-\lambda_0\Delta[g(t_0) + \frac{g''(t_0)}{2!}(t-t_0)^2]+...} d t}.
\end{align}
Let $\alpha = t - t_0$ and $g_0 = g(t_0)$. As long as $t_0$ is in the interval,
\begin{align}
    P(\Delta) &\sim \lambda_0{\int_{-\infty}^\infty [g_0 + \frac{g''_0}{2}\alpha^2]^2 e^{-\lambda_0\Delta[g_0 + \frac{g''_0}{2}\alpha^2]} d \alpha} \\
    &\sim \lambda_0 e^{-\lambda_0\Delta g(0)}{\int_{-\infty}^\infty [g_0 + \frac{g''_0}{2}\alpha^2]^2 e^{\frac{-\lambda_0\Delta g''_0}{2}\alpha^2} d \alpha}.
\end{align}
Let $\mu_0 = \frac{\lambda_0 \Delta}{2},$
\begin{align}
     P(\Delta) &\sim \lambda_0 e^{-2\mu_0 g_0}{\int_{-\infty}^\infty [g_0^{2} + g_0 g''_0 \alpha^{2} + \frac{g''^2_0}{4}\alpha^4] e^{-\mu_0 g''_0\alpha^{2}} d \alpha} \\
     %&=\lambda_0 e^{-2\mu_0 g_0} [\frac{g_0 \sqrt{\pi}}{\mu_0^{0.5} (\frac{g''_0}{2})^{0.5}} + \frac{g_0 2^{1.5} \sqrt{\frac{\pi}{2}}}{\mu_0^{1.5} (\frac{g''_0}{2})^{0.5}} + \frac{g''^2_0 3 \sqrt{\pi}}{16\mu_0^{2.5} (\frac{g''_0}{2})^{2.5}}] \\
     &\sim \lambda_0 e^{-2\mu_0 g_0} [g_0^2 \sqrt{ \frac{\pi} {g''_0} } \mu_0^{-1} + \frac{g_0}{2}\sqrt{ \frac{\pi} {g''_0} } \mu_o^{-1.5} + \frac{3}{16}\sqrt{\frac{\pi}{g''_0}} \mu_0^{-2.5}] \\ 
     &\sim \lambda_0 e^{-2\mu_0 g_0}\sqrt{ \frac{\pi} {g''_0} } [g_0^2 \mu_0^{-1} + \frac{g_0}{2} \mu_o^{-1.5} + \frac{3}{16} \mu_0^{-2.5}]
\end{align}
It can be seen that when the rate vanishes at the minimum ($g_0 = 0$), the power law will  be $-2.5$.
When the rate at the minimum is small there will be a range where the $\mu^{-2.5}$ term dominates.  In this range, we expect
($\Delta  \ll 1/\lambda_0$) and the WTD will have a $-2.5$ power law; otherwise, other power laws may be more applicable. In some processes,  the sinusoidal rate minima might be elevated (nonzero). We derive the analytic solution for an sinusoidal rate function with an elevated minima in the Appendix \ref{elevated_minima}.

\section{Data Analysis}
\begin{figure*}[p] %FIGURE SIZE 7x6 fontsize 12
   \centering
   \includegraphics[width = 13cm]{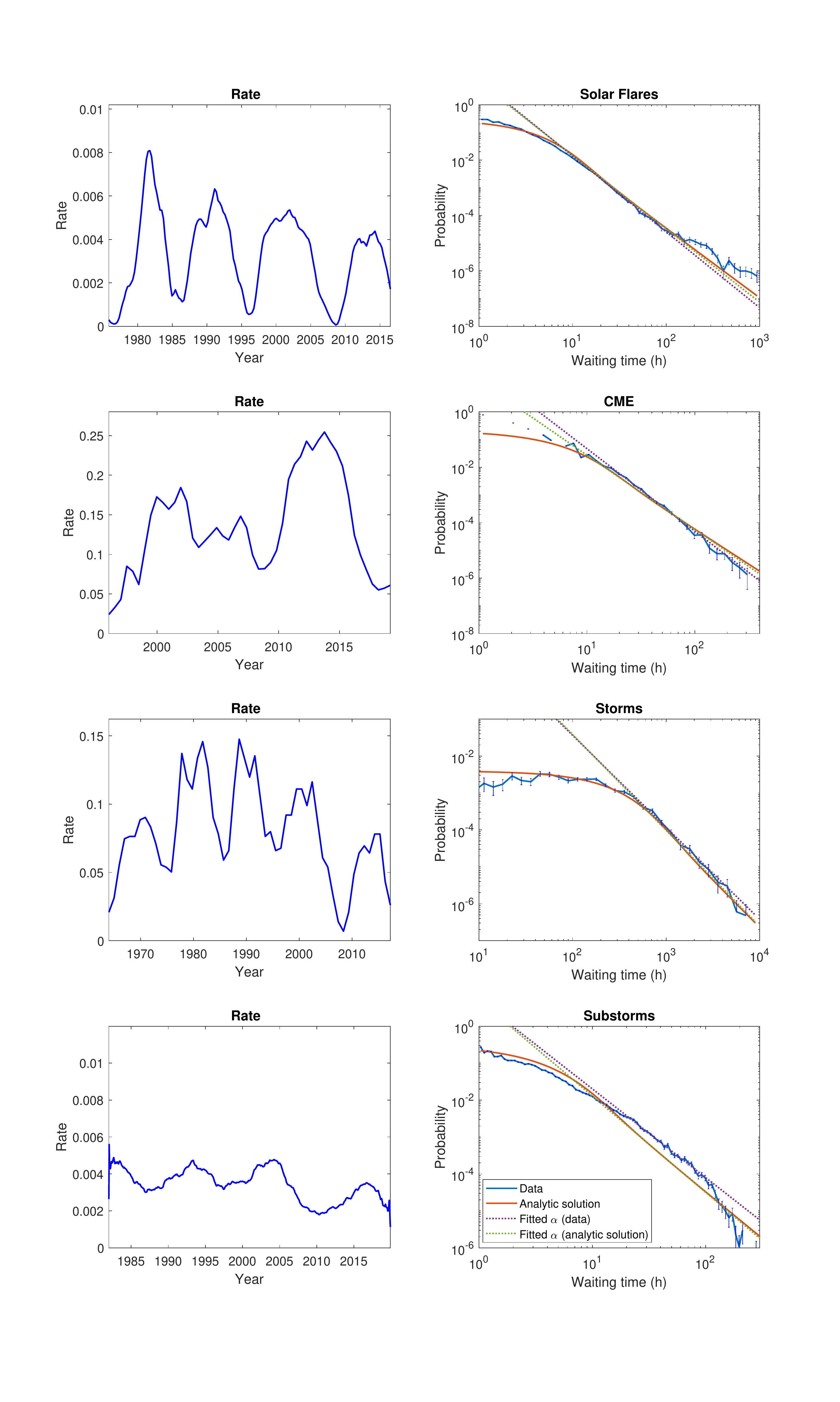}
  \caption{Waiting time probability distribution and rate over the years of the solar, CMEs, Storms and Substorms data. The power law estimate $\alpha$ is over a range as described in table.
              }
         \label{fig:wtd}
\end{figure*}

We present analyses of the waiting time distributions of four different processes: solar flares, CMEs, storms, and substorms. We calculate the waiting time probability distribution using logarithmic binning and plot them alongside the power law fit from equation (\ref{eq:pd}) (see Figure \ref{fig:wtd}). 

\subsection{Solar Flares}
The solar flare data was obtained from the Geostationary Operational Environmental Satellite (GOES) catalog of flares from 1975-2017, available from \url{https://www.ngdc.noaa.gov/stp/solar/solarflares.html}. In keeping with previous studies by \cite{Snelling2020}, we used flares with a minimum peak flux greater than $1.4 \times 10^{-6}$, namely flares of C1 class and above. Event times were set to be the time of the maximum flux. From this sequence consisting of 71,595 flares, we construct a sequence of waiting times. This solar flares waiting time series was the same series analyzed by \cite{Snelling2020}. 

%we calculate the waiting time probability distribution using logarithmic binning and plot them alongside the power law fit using equation (\ref{eq:pd}). 

\subsection{CME}
The Coronal Mass Ejection (CME) list was obtained from the Center for Solar Physics and Space Weather/Naval Research Laboratory (SOHO/LASCO) CME catalog \url{https://cdaw.gsfc.nasa.gov/CME_list/} from 1996 to 2020. We construct a sequence of waiting times from a sequence of 30,321 events.

\subsection{Storms}

The $Dst$ record was obtained from the GSFC/SPDF OMNIWeb interface at \url{https://omniweb.gsfc.nasa.gov}. Using the hourly $Dst$ index between 1963 and 2020, we define the onset of a storm event to be when the index goes below -50 nT and the end of a storm event to be when the index surpasses - 20 nT. The choice of the minimum $Dst$ index is to choose at least moderate storms events (-100 nT $< Dst <$ -50 nT) as characterized in previous study by \cite{Balasis2011}. \cite{Watari1998} also chose $Dst < -50$ nT as a treshold for stroms event. Event times were set to be the onset of the storm. From this sequence of 1276 storm events, we construct a sequence of waiting times.

\subsection{Substorms}
The substorms list was obtained from the SuperMAG substorms list  at \url{http://supermag.jhuapl.edu/mag/}. We created our waiting time sequence from the substorms event times list between 1975 and 2019, which was composed of 68,878 events. 

%\begin{table}
%\begin{tabular}{llccccc}
%& & $\alpha$& $\Delta_{min} (hours)$ & $\Delta_{max}$ (hours)& $\chi^2$& $p$\\
%\hline
%Solar Flares& Data& $-2.74 \pm 0.06$& $19.6$& $102.6$& 0.42& $1.7 \times 10^{-8}$\\
%& Analytic Solution& $-2.65 \pm 0.05$& $19.6$& $102.6$& 0.038& $3.4 \times 10^{-15}$\\
%CMEs& Data& $-2.98 \pm 0.05$& $19.8$& $189.0$& 1.03& $3.2 \times 10^{-7}$\\
%& Analytic Solution& $-2.68 \pm 0.05$& $19.8$& $189.0$& 0.098& $1.01 \times 10^{-14}$\\
%Storms& Data& $-2.49 \pm 0.15$& $566.1$& $4496.472$& 0.38& $1.8 \times 10^{-6}$ \\
%& Analytic Solution& $-2.60 \pm 0.2$& $566.1$& $4496.472$& 0.053& $1.0 \times 10^{-10}$\\
%Substorms& Data& $-2.40 \pm 0.06$& $24.9$& $98.9$& 1.1& $8.5 \times 10^{-11}$\\
%& Analytic Solution& $-2.61 \pm 0.13$& $24.9$& $98.9$ &0.0048& $2.8 \times 10^{-35}$
%\end{tabular}
%\caption{\label{tab:table2} Linear least squares fit for the WTDs of data and analytic solution, eq. %\ref{eq:pd}, where the fitted power law is for $ \Delta_{min} < \Delta < \Delta_{max}$.} 
%\end{table}

\begin{table}
\begin{tabular}{llccccc}
& & $\alpha$& $\Delta_{min} (hours)$ & $\Delta_{max}$ (hours)& $\chi^2$& $p$\\
\hline
Solar Flares& Data& $-2.74 \pm 0.06$& $19.6$& $102.6$& 0.42& $\ll 0.001$\\
& Analytic Solution& $-2.65 \pm 0.05$& $19.6$& $102.6$& 0.038& $\ll0.001$\\
CMEs& Data& $-2.98 \pm 0.05$& $19.8$& $189.0$& 1.03& $\ll0.001$\\
& Analytic Solution& $-2.68 \pm 0.05$& $19.8$& $189.0$& 0.098& $\ll0.001$\\
Storms& Data& $-2.49 \pm 0.15$& $566.1$& $4496.472$& 0.38& $\ll0.001$\\
& Analytic Solution& $-2.60 \pm 0.2$& $566.1$& $4496.472$& 0.053&$\ll0.001$\\
Substorms& Data& $-2.40 \pm 0.06$& $24.9$& $98.9$& 1.1&$\ll0.001$\\
& Analytic Solution& $-2.61 \pm 0.13$& $24.9$& $98.9$ &0.0048& $\ll0.001$
\end{tabular}
\caption{\label{tab:table2} Linear least squares fit for the WTDs of data and analytic solution, eq. \ref{eq:pd}, where the fitted power law is for $ \Delta_{min} < \Delta < \Delta_{max}$.} 
\end{table}

\section{Results and Discussions}
Figure \ref{fig:wtd} plots waiting time distribution for solar flares, CMEs, storms, and substorms. The figure shows the existence of power laws for  $ \Delta_{min} < \Delta < \Delta_{max}$ region as described in Table \ref{tab:table2}. The power law distribution does not fit the actual WTD for the entire tail domain of the datasets perhaps due to some underlying dynamics that compete with the cyclical behavior. For the solar flares, it seems that the WTD at long waiting times may have a shallower slope. It may be the case that during solar minima, there are impulsive events which have been associated with shallower power law slope \cite{Aschwanden2010}. The power law slope of the CMEs is somewhat steeper than the WTD derived from the sinusoidal rate.  Steeper slopes can be found when the minima of the sinusoidal rate is elevated (See Appendix \ref{elevated_minima}). If analyzed segment by segment, \cite{Wheatland2003} found a power law $\alpha \sim -2.36 \pm 0.11$ for CMEs between 1996-2001, the period of lowest CMEs activity but found a power law $\alpha \sim -2.98 \pm 0.20$ for the years of 1999-2001, a period of higher activity where the minima of the rate function is elevated.  These power laws are reasonably similar to the analytic distribution. However, there is a small difference, suggesting that additional processes should be considered beyond simple sinusoidal driving.  For the substorms, the internal magnetospheric dynamics likely affects the distribution at shorter waiting times, thus the power law only matches at long waiting time. The sudden steepness of the slope may be attributed to the elevated minima of the rate function (see Appendix \ref{elevated_minima}). The power law distribution for the storms fits much better than for the substorms, indicating that the cyclical behavior of the external driver is generally more important. In other words, there is less internal dynamics involved than for substorms.  %\cite{Michelis2011} suggested that a sequence of magnetospheric substorms may drive moderate storms and very large storms may drive magnetospheric substorms. 

Although this paper focuses mostly on events impacted by the periodicity of the solar activity, the waiting time distribution of any nonstationary Poisson process that has a sinusoidal rate with a minimum near zero will have a power law of $-2.5$ at large waiting times. Even more generally, for any nonstationary Poisson processes that has a continuous rate with a minimum close to zero, the WTD will have a power law of $-2.5$ at large waiting times as long as the minimum rate is small. Although, natural processes might not reach that asymptotic point due to the rarity of the events at large waiting time and potential data gaps. Also, for processes with an elevated minima, the slope at large waiting times may steepens dramatically and loses the power law characteristic. It may be inferred that, for longer waiting times, the internal dynamics may be a lot less important than what is driving the system itself. In this case, it is how the minima behaves.

The power laws of  WTDs are often thought to result from SOC processes or turbulent interactions.  In this work we have shown that simple periodic driving of a system with a random response may also produce a power law of -2.5.  Because the solar magnetic activity cycle is a primary driver of dynamics throughout the solar system, it is not surprising that such power laws are seen in solar flares, CME, storm and substorm datasets.  It is also to be noted that because long waiting times primarily occur when the rate is minimum, that the tail of power laws provide information about conditions at the minima in activity cycles, and may provide insight as to the underlying dynamics at solar minima and the overlap between cycles.

This work is supported by NASA grants 
NNX15AJ01G, NNH15AB17I, NNX16AQ87G, 80NSSC19K0270, 80NSSC19K0843, 80NSSC18K0835, 80NSSC20K0355, NNX17AI50G, NNX17AI47G, 80HQTR18T0066, 80NSSC20K0704 and NSF grants AGS1832207 and AGS1602855 and Andrews University FRG 201119.
%%BIBLIOGRAPHY
\bibliographystyle{unsrt}
\bibliography{bibliography.bib}
\newpage

\appendix
\section{Sinusoidal Rate Function with Elevated Minimum, $\lambda_{min}>0$}\label{elevated_minima}

For the nonstationary Poisson process we have approximately:

\begin{align}
P(\Delta )=\frac{\int_0^T \lambda (t)^2 e^{-\lambda (t)\Delta } dt}{\int_0^T \lambda (t)dt}.
\end{align}

Suppose we choose a sinusoidal dependence and consider a period (it will be the same result for multiple periods):

\begin{align}
\lambda (t)=\lambda_0 \left(1+\frac{\beta }{2}\cos \omega t\right)
\end{align}

where 

\begin{align}
\delta \lambda =\lambda_{max} -\lambda_{min} =\beta \lambda_0
\end{align}
For which 
\begin{align}
\int_0^T \lambda (t)dt=\lambda_0 T
\end{align}
when integrated over a period.

Then we have

\begin{align}
    P(\Delta )=\frac{\lambda_0 e^{-\lambda_0 \Delta } }{T}\int_0^T (1+\frac{\beta }{2}\cos \omega t)^2 e^{-\frac{\beta \lambda_0 \Delta }{2}\cos (\omega t)} dt.
\end{align}

Changing variables of integration:

$$\theta =\omega t$$ 

with

$$d\theta =\omega dt$$

$$P(\Delta )=\frac{\lambda_0 e^{-\lambda_0 \Delta } }{\omega T}\int_0^{\omega T} (1+\frac{\beta }{2}\cos (\theta ))^2 e^{-b\cos (\theta )\Delta } d\theta$$

where $b=\frac{\lambda_0 \beta \Delta }{2}$.

Because this is a periodic signal, the statistics can all be obtained by considering one period. Let us consider $P(\Delta )$ determined over a period $\omega T=2\pi.$ Then

\begin{align}
    P(\Delta )=\frac{\lambda_0 e^{-\lambda_0 \Delta } }{2\pi }\int_0^{2\pi } (1+\frac{\beta }{2}\cos \theta )^2 e^{-b\cos \theta } d\theta. \label{eq:int_mod}
\end{align}

The integral can be performed using the Bessel function identity (\cite{Abramowitz1972} - 9.6.34)
\begin{align}
    e^{z\cos \theta } =\sum_{n=-\infty }^{\infty } I_n (z)e^{in\theta }
\end{align}
so that  
\begin{align}
    e^{-b\cos \theta } =\sum_{n=-\infty }^{\infty } I_n (-b)e^{in\theta } =\sum_{n=-\infty }^{\infty } (-1)^n I_n (b)e^{in\theta }
\end{align}

where 
\begin{align}
    I_n (-z)=(-1)^n I_n (z)  \text{(\cite{Abramowitz1972} - 9.6.30)}.
\end{align}
Then

\begin{align}
P(\Delta ) &=\lambda_0 e^{-\lambda_0 \Delta } \sum_{n=-\infty }^{\infty } I_n (b)\frac{1}{2\pi }\int_0^{2\pi } e^{in\theta } {\left(1+\frac{\beta }{2}\left(\frac{e^{i\theta } +e^{-i\theta } }{2}\right)\right)}^2 d\theta \\
&=\lambda_0 e^{-\lambda_0 \Delta } \sum_{n=-\infty }^{\infty } (-1)^n I_n (b)\left\lbrack \frac{1}{2\pi }\int_0^{2\pi } e^{in\theta } \left(1+\frac{\beta }{2}\left(e^{i\theta } +e^{-i\theta } \right)+\frac{\beta^2 }{4}\left(\frac{e^{2i\theta } +2+e^{-2i\theta } }{4}\right)\right)d\theta \right\rbrack \\
&=\lambda_0 e^{-\lambda_0 \Delta } \sum_{n=-\infty }^{\infty } (-1)^n I_n (b)\left\lbrack \left(1+\frac{\beta^2 }{8}\right)\delta_{n,0} +\frac{\beta }{2}\left(\delta_{n,-1} +\delta_{n,1} \right)+\frac{\beta^2 }{16}(\delta_{n,-2} +\delta_{n,2} )\right\rbrack \\
&=\lambda_0 e^{-\lambda_0 \Delta } \left\lbrack \left(1+\frac{\beta^2 }{8}\right)I_0 (b)-\frac{\beta }{2}\left(I_{-1} (b)+I_1 (b)\right)+\frac{\beta^2 }{16}(I_{-2} (b)+I_2 (b))\right\rbrack
\end{align}
and making use of the property
\begin{align}
    I_{-n} (z)=I_n (z), \text{(\cite{Abramowitz1972} - 9.6.6),}
\end{align}
 we have

\begin{align}
    P(\Delta ) &=\lambda_0 e^{-\lambda_0 \Delta } \left\lbrack \left(1+\frac{\beta^2 }{8}\right)I_0 (b)-\beta I_1 (b)+\frac{\beta^2 }{8}I_2 (b)\right\rbrack\\
P(\Delta ) &=\lambda_0 e^{-\lambda_0 \Delta (1-\frac{\beta }{2})} e^{-b} \left\lbrack \left(1+\frac{\beta^2 }{8}\right)I_0 (b)-\beta I_1 (b)+\frac{\beta^2 }{8}I_2 (b)\right\rbrack
\end{align}

\begin{figure*} 
   \centering
   \includegraphics[width = 13cm]{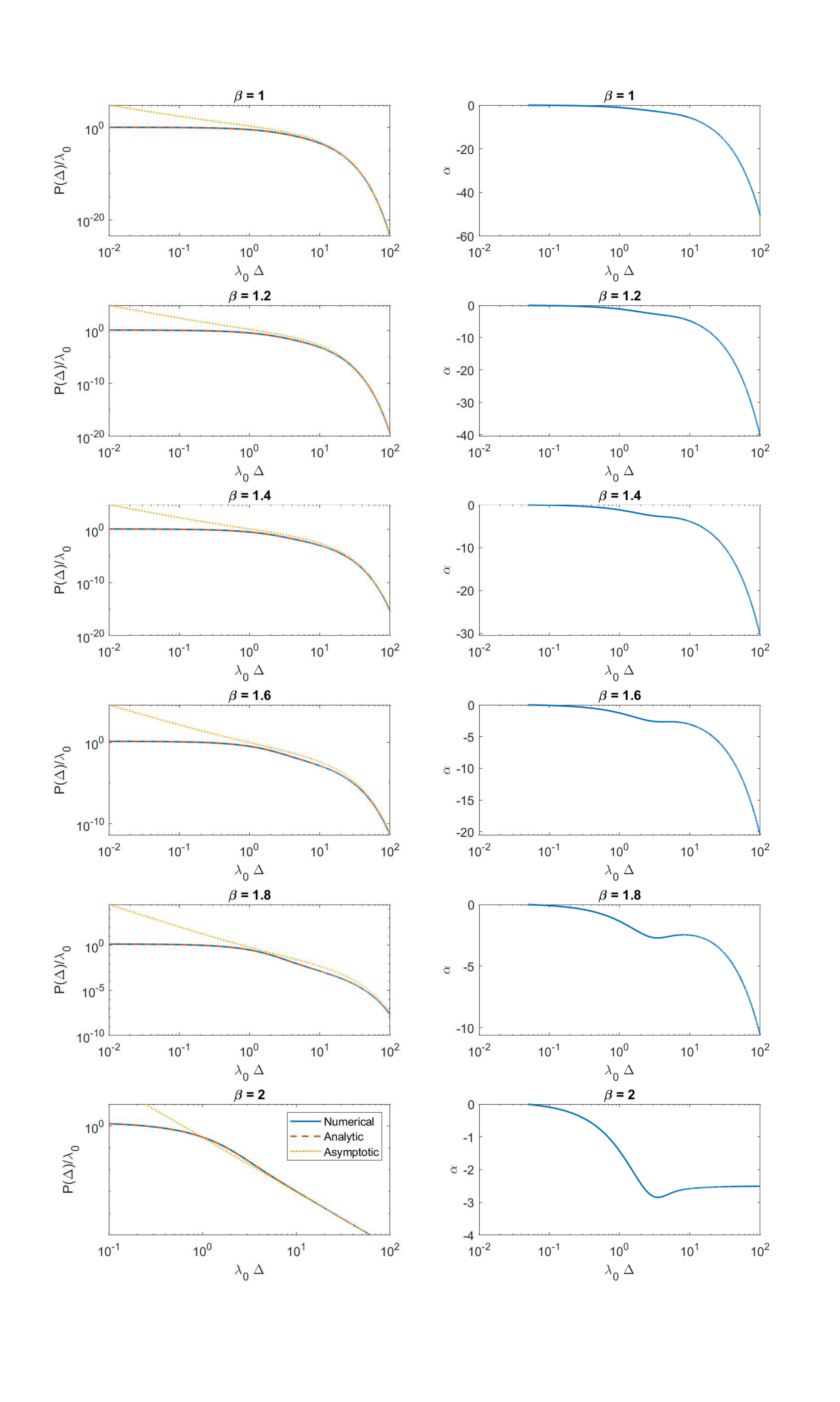}
  \caption{The left panels are the waiting time distributions with numerical and analytic solutions to eq. (\ref{eq:int_mod}) corresponding to various $\beta$ values. The right panels are the "local" power law estimation.
              }
         \label{fig:sim_mod}
\end{figure*}

Now we perform the asymptotic analysis. For large $\Delta$ we have 
\begin{align}
    I_n (z)\sim \frac{e^z }{\sqrt{2\pi z}}\left\lbrack 1-\frac{\mu -1}{8z}+\frac{(\mu -1)(\mu -9)}{2!(8z)^2 }-\frac{(\mu -1)(\mu -9)(\mu -25)}{3!(8z)^3 }+...\right\rbrack
\end{align}

where $\mu =4n^2$. And so
\begin{align}
    I_0 (z)e^{-z} &\sim \frac{1}{\sqrt{2\pi z}}\left\lbrack 1+\frac{1}{8z}+\frac{9}{128z^2 }+...\right\rbrack\\
    I_1 (z)e^{-z} &\sim \frac{1}{\sqrt{2\pi z}}\left\lbrack 1-\frac{3}{8z}-\frac{15}{128z^2 }+...\right\rbrack\\
    I_2 (z)e^{-z} &\sim \frac{1}{\sqrt{2\pi z}}\left\lbrack 1-\frac{15}{8z}+\frac{105}{128z^2 }+...\right\rbrack.
\end{align}
Therefore,
\begin{align}
    P(\Delta )\sim \lambda_0 e^{-\lambda_0 \Delta (1-\frac{\beta }{2})} \frac{1}{\sqrt{2\pi b}}\left\lbrack {\left(1-\frac{\beta }{2}\right)}^2 +\left(1+\frac{7\beta }{2}\right)\left(1-\frac{\beta }{2}\right)\left(\frac{1}{8b}\right)+\frac{57\beta^2 +60\beta +36}{512b^2 }+...\right\rbrack
\end{align}
The power law depends on the choice of $\beta$.  In the case that $\beta =2$ we recover $P\sim b^{-2.5}$, otherwise it drops off exponentially. In Fig. \ref{fig:sim_mod} we plot the WTDs and corresponding "local" power law estimation for various values of $\beta$.

\end{document}